# A novel multi-GPU parallelization paradigm for SPH applied to solid mechanics in complex industrial applications

Thomas Unfer, Anthony Collé and Jérôme Limido

IMPETUS AFEA
Plaisance du Touch, France
thomas@impetus.fr

*Abstract*—A novel parallelization paradigm has been developed for multi-GPU architectures. Classical multi-GPU parallelization for SPH rely on domain decomposition. In our approach each particle can be assigned to a GPU independently of its position in space. This ensures a kind of natural load balancing because the number of particles per GPU remains constant. The data exchange domain is no more a surface as in the classical approach or in mesh-based method, but it is a volume which is growing as mixing occurs in time between the particles assigned to different GPUs. This growth must be prevented because the efficiency in terms of computation time and memory consumption is rapidly dropping with the mixing. A simple heuristic is suggested to periodically detect particles to swap between GPUs in order to keep the exchange volume close to a surface. The final decomposition is much alike a domain decomposition and can be applied to any geometry. We developed this approach within the framework of the industrial code IMPETUS Solver® using the Gamma-SPH-ALE method. Industrial applications in the space and military fields are presented. They demonstrate the relevance of the approach developed in an industrial context and for complex applications.

## I. Introduction

One of the key points of the implementation of the SPH method is to efficiently find the interacting particles laying within a particle kernel function support. A popular algorithm for GPU based simulation is the bucket sort or radix sort algorithm. At the beginning of each iteration, the computational domain is divided into square or cubic buckets of size 2h with h the maximum smoothing length of the particles. Each particle is given a hash number corresponding to the bucket in which it is located. The particles are then sorted with respect to this hash number by the radix sort algorithm. The search for the neighbours of the particle can then be restricted to its bucket and to the surrounding buckets. This algorithm is particularly efficient on a single GPU. For multi-GPU architecture, the common practice is to use a domain decomposition and then exchange data near the subdomain boundaries (See [1], [2]). The domain decomposition paradigm implies to tackle two difficult questions: first, an initial domain decomposition is to be determined. Second, if the solution is strongly time varying, dynamic load balancing shall be implemented to keep the workload balanced over the GPUs all along the simulation. This paper explores an alternate pathway: particle decomposition, a method specially intended for particle methods. The intended target hardware configuration is a typical workstation equipped with a reasonably large number of GPUs (≤16). The first step consists in developing an algorithm that can handle multi-GPU computations with particles located at any position within the computational domain. The second step consists in developing a particle exchange algorithm whose purpose is to minimize the amount of data to be exchanged from GPU to GPU. The SPH method used in all applications of this paper is the Gamma-SPH-ALE method of Collé et al [3] implemented in the industrial code IMPETUS Solver®.

## II. From domain decomposition to particle decomposition

The particle decomposition algorithm has been developed considering that the computational workload is directly proportional to the particle number. This is true for the SPH method provided that all particles have roughly the same number of neighbors which means that all particles shall have the same smoothing length. Under this assumption the computation is well balanced if all GPUs have the same number of particles. However, the communication effort may be far from optimal. The algorithm uses an additional hash number for multi-GPUs communications. If particles from other GPUs are present in the particle's bucket or in the surrounding buckets, the communication hash number is equal to the hash number, otherwise it is equal to some garbage collector hash. Then the particles can be sorted according to this communication hash, so that the data that shall be broadcasted to the other GPUs can be isolated. Those particles will be called "MGPU boundary particles" in the sequel. The algorithm is as follows:
  a. Each GPU computes its particle hash numbers.
  b. The hash numbers are broadcasted to all GPUs.

c. Each GPU computes its particle communication hash numbers.
d. The computational data from all the MGPU boundary particles are broadcasted to the other GPUs.
e. In the meantime, each GPU can compute its local particle interaction routines.
f. Once the data is received, each GPU can compute its local particle interaction with the MGPU boundary particles.
g. Each GPU can integrate its equations.

Note: if the SPH formulation implies to compute several interaction loops for one iteration, steps d. to f. are duplicated for each loop.

Figure 1 shows an example of the particle decomposition for a 2D Dam break computed on 3 GPUs. The initial decomposition is much alike an optimal domain decomposition, but once the wave reaches the opposite wall and strikes back the particles start to mix. Nevertheless, the computation proceeds nicely. The MGPU boundary particles, which are located onto two surfaces at the initial stages, are progressively spreading in the whole computation domain. The communication load increases accordingly and will eventually slow down the computation. The computational workload remains balanced because the particle number in each GPU is the same. So, it is possible to try to minimize the communication workload by swapping particles from a GPU to the other. The computational workload will stay balanced for all the swapping operations. A simple heuristic to achieve this goal will be introduced in the next section.

using radix sort. This algorithm is intended to be activated periodically. A sensible value out of numerical experiments is to activate the balancing algorithm every 20 integration loops.

A. *Single particle type*

If the calculation consists of a single type SPH particles with constant smoothing length, the following algorithm can be applied. Every GPU computes its own particle cloud centre of gravity (Cg). A balancing hash number is defined based on the projection of the particle position on the line from one GPU Cg to the other. If the projection is located on the own GPU side of the middle of the segment of the two Cgs, the particle is disregarded for swapping (hash number 0). If the projection is located beyond the other Cg, the particle is given the maximum swapping hash number (typically 1000). In between the balancing hash number is proportional to the distance between the middle of the Cgs and the other Cg up to the maximum swapping hash number. If both GPUs have particles with positive hash numbers, some particle swapping occurs. The number of particles to be swapped is determined by the minimum number of particles with a positive hash number on each GPU. The particle data can be sorted by the balancing hash number and the corresponding particles can be swapped between the two GPUs. The algorithm is then repeated for each GPU pair.

Figure 2 shows the same Dam break test case as shown on Figure 1 with enabled dynamic load balancing. The resulting particle decomposition is much alike a domain decomposition except during some transient phases, but the corresponding particle clouds remain compact all the time. The information exchange subset is always close to a surface and never exceeds a small percentual of the global particle number.

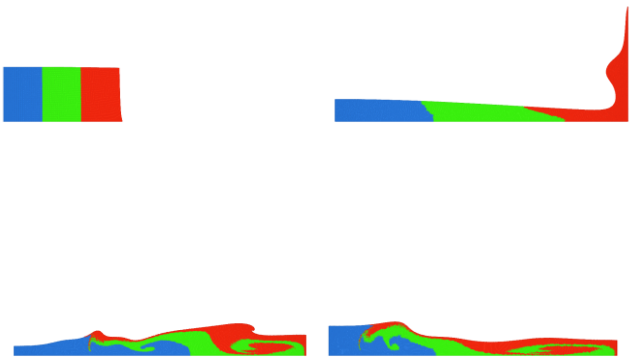

Figure 1.    Particle decomposition over 3 GPUs for a 2D Dambreak without dynamic load balancing

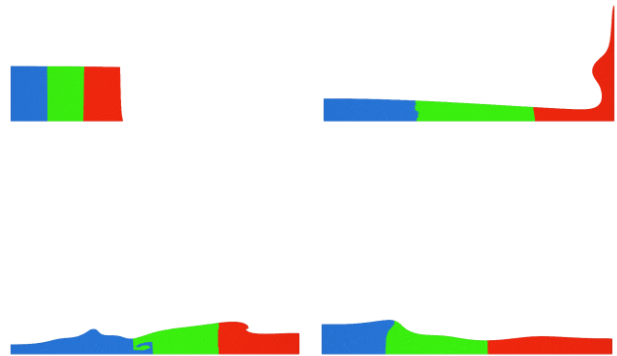

Figure 2.    Particle decomposition over 3 GPUs for a 2D Dambreak with dynamic load balancing

III. A HEURISTIC FOR COMMUNICATION LOAD BALANCING

The goal of this section is to define an on-the-fly algorithm that can be used to keep the communication load for particle decomposition close to the one that would be used with domain decomposition. The idea is to use consecutive particle swapping between GPU pairs based on particle hashing and

B. *Multiple particle types*

In solid mechanics applications, multiple particle types with different smoothing lengths and different behaviours are often present, particles can be eroded (taken away from the computation). In general, it is impossible to evaluate a priori the computation cost of each particle part. A practical solution is to consider that all particle types shall be balanced over each GPU to equilibrate the computational workload. The swapping algorithm of section III.A can then be applied for each particle type. However, this may lead to an independent decomposition from particle type to particle type. To overcome this to a certain extent, a biased Cg' can be used. The idea is to shift the Cg used for balancing from the actual Cg of the particles of the desired type towards the global Cg of the particles:

$$Cg'_{typeA} = \alpha Cg_{alltypes} + (1-\alpha) Cg_{typeA} \text{ with } \alpha = 0.1$$

## IV. APPLICATIONS TO SOLID MECHANICS IN THE SPACE AND MILITARY FIELD

### A. HyperVelocity Impact (HVI) of an Aluminium bead on a Aluminium plate

The increase of the number of space debris in low Earth orbit is a growing concern for the space industry. The modelling of the impact of such small debris on a satellite metallic envelope is becoming a key issue and the SPH method is the ideal candidate to treat such problems. Figure 3 shows a comparison between experiment (up) and simulation (down) of the impact of an aluminium bead of 1.5 mm radius on a 2mm thick aluminium plate at 4.05 km/s with an impact angle tilted of 32° with respect to the normal direction to the plate. The material model used for aluminium is Johnson-Cook's constitutive model in the ideal elastic perfectly plastic limit with a Mie-Gruneisen equation of state and a cut-off pressure to consider the spalling mechanism of aluminium. No additional damage modelling is introduced into the model.

The simulation was run with 6.5 million of particles on one, two and three GPUs (Nvidia Titan Xp). The corresponding particle decomposition is shown on figure 4 in the multi-GPU cases at the beginning and at the end of the simulation. It can be noticed that the fragment cloud generated in the multi-GPU cases are slightly different from the one in the mono-GPU case. Those discrepancies can be explained by the highly non-linear equations solved and by the fact that the arithmetic operations are not performed in the very same order depending on the GPU number. Still the clouds are statistically equivalent. The runs lasted respectively 2h36mn, 1h25mn and 1h6mn, giving a strong scaling speed-up of 1.84 in the 2 GPU case and 2.36 in the 3 GPU case.

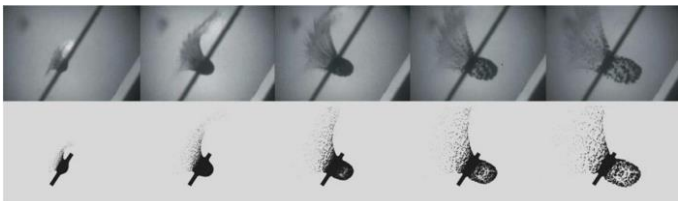

Figure 3.    Simulation vs Experiment (Courtesy of Thiot Engineering) of a hypervelocity impact on an aluminium plate

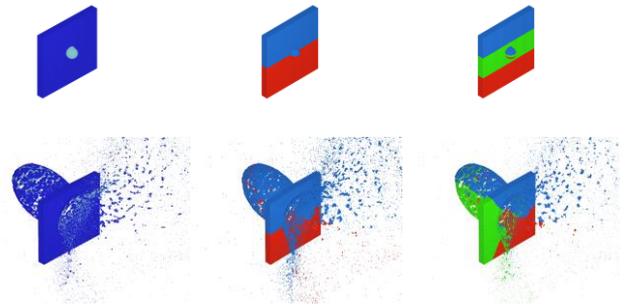

Figure 4.    Particle decomposition on 2 and 3 GPUs for the HVI case (mono GPU run on the left, dark blue: plate, light blue: bead)

### B. Taylor bar impact on a symmetry plane

Taylor bar impacts are classical experiments to investigate material failure under fast loading. It consists of the impact of a metallic cylinder hitting a hard target at high velocity. In this section a 200g steel rod is considered. It is 8cm long and has a diameter of 2 cm. The rod is hitting a symmetry plane at 300 m/s. The constitutive model for ductile metal of the IMPETUS solver is used, it accounts for thermal softening and strain hardening. Some initial random damage is present in the rod and the Cockcroft-Latham failure criterion is used. When a particle is completely damaged, the deviatoric part of its stress tensor is removed and the particle recovers a fluid behaviour.

This simulation was done on two Nvidia RTX 8000 GPUs with 58 million particles in 3 days. The rendering has been done using the Sheppard sum with the SPH volume interpolator of the Paraview software. The results are presented on figures 5 to 8. The model shows the propagation of the cracks through adiabatic shear layers and the formation of a spiral shear pattern. This fracture pattern has already been identified in [4] and confirmed by Finite Element simulations. The SPH solution is well in line with the experimental data taken from [4] and recalled on figure 9.

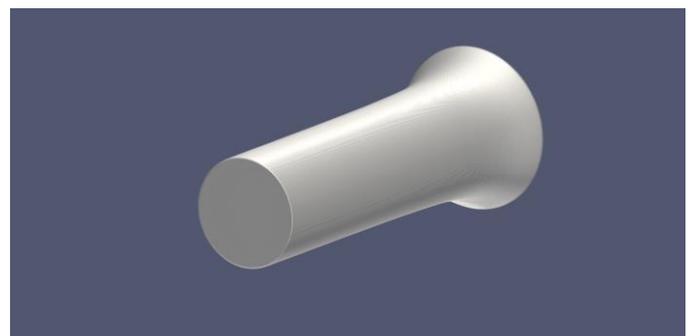

Figure 5.    Taylor impact: solution at 54 μs

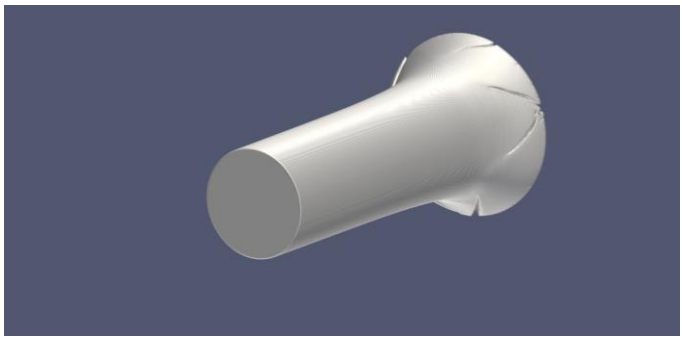

Figure 6.　　　　Taylor impact: solution at 90 µs

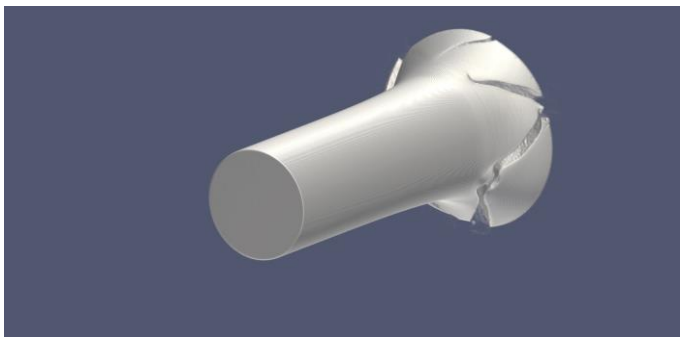

Figure 7.　　　　Taylor impact: solution at 180 µs

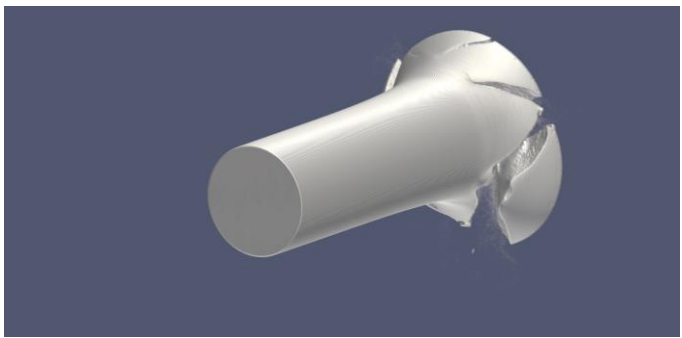

Figure 8.　　　　Taylor impact: solution at 360 µs

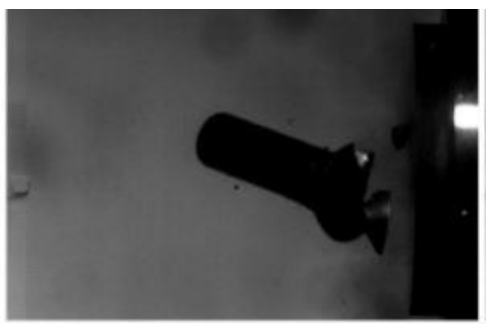

Figure 9.　　　　Experiment taken from Ragvak et al[4]

### C. Taylor bar impact on a encased ceramic target

This section presents another example of application. This time an encased ceramic target is hit by the same steel rod as in section IV.B. The impact velocity is 900 m/s and the domain size has been reduced by the use of a symmetry plan. The simulation was done with 85 million particles and took 5 days. The ceramic is modelled with a Johnson-Holmquist model for brittle material. The visualization is done using Paraview SPH volume interpolator and Sheppard summation as in the previous section. The first instants after the impact are shown on figures 10 to 12, the propagation of the shock wave through the ceramic is clearly visible. It reaches the back end of the ceramic domain and bounces forth.

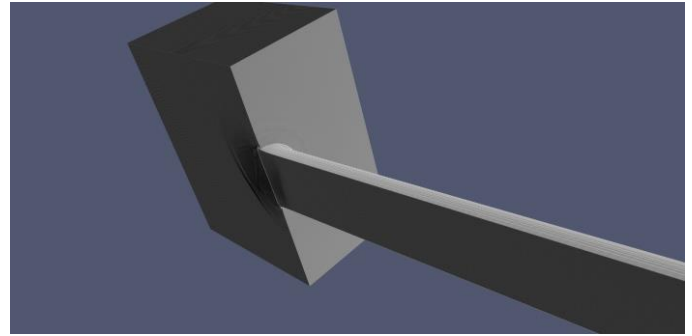

Figure 10.　　　　Taylor impact on ceramic: solution at 1.2 µs

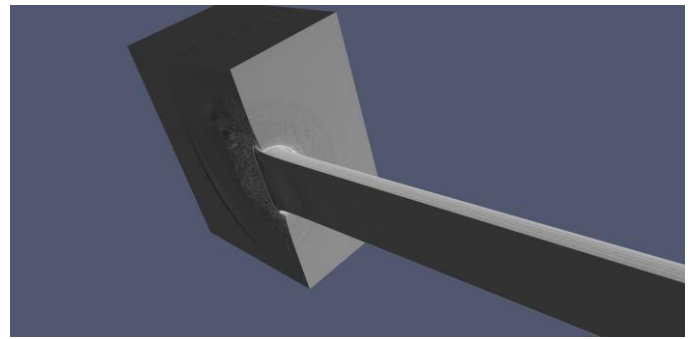

Figure 11.　　　　Taylor impact on ceramic: solution at 3.6 µs

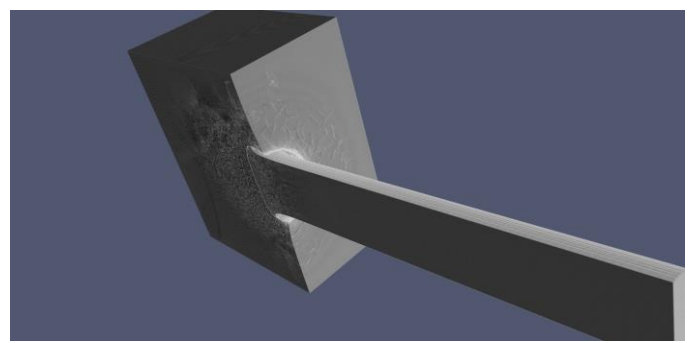

Figure 12.　　　　Taylor impact on ceramic: solution at 6.0 µs

The evolution on the longer term is shown on figures 13 to 15. The cracks propagation patterns of the ceramic material appear clearly on figure 13. Once the ceramic is broken, fragments are expelled outwards. The only possible way for the ceramic fragments is outwards because the boundary conditions (encasement) make it impossible for the fragments to move sideways. The deformation pattern of the rod is also quite different from the previous section because the impact velocity is higher. The rod is also contained by the surrounding ceramic material.

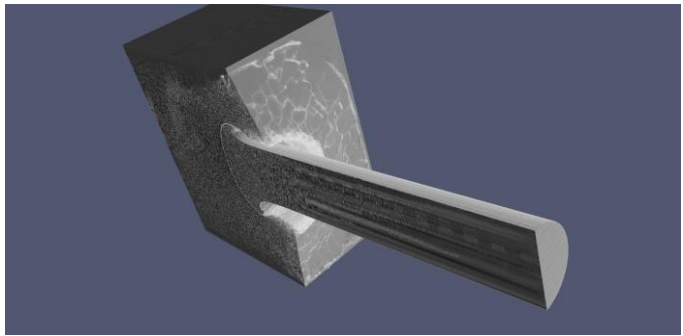

Figure 13.　　　Taylor impact on ceramic: solution at 24 µs

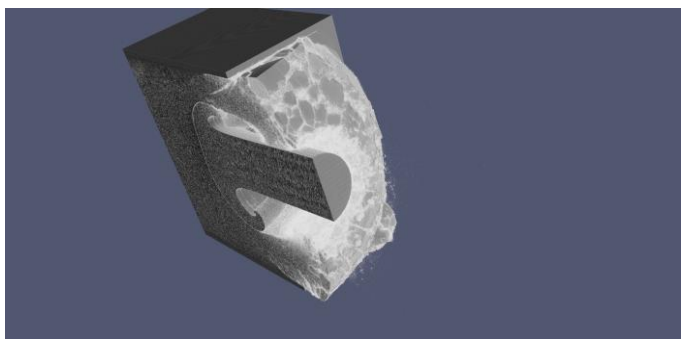

Figure 14.　　　Taylor impact on ceramic: solution at 60 µs

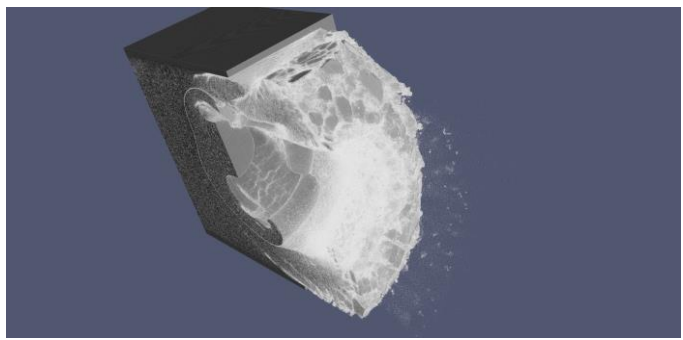

Figure 15.　　　Taylor impact on ceramic: solution at 120 µs

## V. CONCLUSION AND PERSPECTIVES

A new multi-GPU parallelization paradigm has been introduced, based on particle decomposition rather than domain decomposition. The computational workload is balanced by construction in this approach and the communication workload is kept to a reasonably low level thanks to a simple heuristic. The method has been successfully applied to solid mechanics. This approach could be generalized to SPH simulations with varying particle sizes thanks to the introduction of a cost function considering the particle actual neighbour number in order to estimate the computational workload. Another perspective is to generalize the approach to massively parallel GPU architectures.